\documentclass{article}
\usepackage{times}
\usepackage{amsfonts}
\usepackage[pdfmark,colorlinks]{hyperref}
\begin{document}
\title{Noncommutative Manifolds from the Higgs Sector of Coincident D--Branes}
\author{Jos\'e M. Isidro\\
Instituto de F\'{\i}sica Corpuscular (CSIC--UVEG)\\
Apartado de Correos 22085, Valencia 46071, Spain\\
{\tt jmisidro@ific.uv.es}}

\maketitle

\begin{abstract}

The Higgs sector of the low--energy physics of $n$ of coincident D--branes contains the necessary elements for constructing noncommutative manifolds. The coordinates orthogonal to the coincident branes, as well as their conjugate momenta, take values in the Lie algebra of the gauge group living inside the brane stack. In the limit when $n\to\infty$ (and in the absence of orientifolds), this is the unitary Lie algebra $u(\infty)$. Placing a smooth manifold ${\cal K}$ orthogonally to the stack of coincident D--branes one can construct a noncommutative $C^{\star}$--algebra that provides a natural definition of a noncommutative partner for the manifold ${\cal K}$.

\end{abstract}

\tableofcontents

\section{Introduction}\label{ramalloguarrolavatelospies}

The effective action describing the low--energy physics of $n$ parallel, coincident D$p$--branes contains a Yang--Mills sector and a Higgs sector. 
It has been observed in ref. \cite{WITTEN} that the latter provides us with a dynamical system whose classical phase space requires the notion of Lie--algebra valued coordinates and momenta. This letter, a followup to ref. \cite{GOLM}, is devoted to analysing the classical and quantum dynamics of Lie--algebra valued coordinates and momenta as given by the Higgs sector.

Specifically, the superposition of $n$ such D$p$--branes produces a $u(n)$ gauge theory on their common $(p+1)$--dimensional worldvolume. Although this theory is supersymmetric, we will concentrate throughout on its bosonic sector. Now ${u}(n)={u}(1)\times {su}(n)$ is not simple, but separating out the centre--of--mass motion we are left with the simple Lie algebra ${su}(n)$. (The gauge groups $so(n)$ and $sp(n)$ can be obtained by adding orientifolds to the stack of $n$ coincident branes as done, {\it e.g.}, in ref. \cite{THEISEN}).
Let $A_{\mu}$ be an ${su}(n)$--valued gauge field on the D--brane stack, and let us separate its components into longitudinal and transverse parts to the D--branes, $A_{\mu}=(A_l, A_t)$. Longitudinal components $A_l$ are adjoint--valued ${su}(n)$ matrices, {\it i.e.}, Yang--Mills gauge fields. 
Transverse components $A_t$ describe D--brane fluctuations that are orthogonal to the D--branes, {\it i.e.}, Higgs fields. They are thus identified with transverse coordinates, so they are more properly denoted $x_l$ instead of $A_l$. Modulo numerical factors, the bosonic part of super Yang--Mills theory dimensionally reduced to $p+1$ dimensions is
\begin{equation}
S_{\rm YM}^{(p+1)}=\int{\rm d}^{p+1}\xi\,{\rm tr}\,({\cal F}_{ll'}^2+2{\cal F}^2_{lt}+{\cal F}^2_{tt'}),
\label{labastidachupamelapollamarikondemierda}
\end{equation}
where $l,l'$ ($t,t'$) are longitudinal (transverse) indices. D--boundary conditions remove all derivatives in the $t$ directions, and (again up to numerical factors) eqn. (\ref{labastidachupamelapollamarikondemierda}) becomes
\begin{equation}
S_{\rm YM}^{(p+1)}=\int{\rm d}^{p+1}\xi\,{\rm tr}\,{\cal F}_{ll'}^2 - \int{\rm d}^{p+1}\xi\,{\rm tr}\,\left(\frac{1}{2} (D_lx^t)^2-\frac{1}{4}[x^t,x^{t'}]^2
\right),
\label{cesargomezquetefollenkabrondemierda}
\end{equation}
where $D_lx^t=\partial_lx^t+{\rm i}[A_l, x^t]$ is the longitudinal, gauge--covariant derivative of transverse coordinates $x^t$. The appearance of matrix--valued coordinate functions can be motivated in the relation of D$p$--branes to Chan--Paton factors via T--duality. When $p=0$ we have the important case of the M(atrix) model of M--theory in the light--cone gauge \cite{BFSS}, where the limit $n\to\infty$ is taken.

We are interested in the  Higgs sector. It is described by the {\it transverse}\/ coordinates to the D--brane, given by the following terms in eqn. (\ref{cesargomezquetefollenkabrondemierda}):
\begin{equation}
S_{\rm Higgs}=- \frac{1}{2}\int{\rm d}^{p+1}\xi\,{\rm tr}\,\left(\frac{1}{2} (D_lx^j)(D_lx^j)-\frac{1}{2}\sum_{i\neq j}[x^i,x^{j}]^2.
\right),
\label{luisibanezquetefollenkabrondemierda}
\end{equation}
Above, $l=0, 1,, \ldots p$ runs over the longitudinal coordinates to the D--brane, and $i,j=p+1, \ldots, d$ run over the transverse coordinates to the D--brane. The latter is embedded within $d$--dimensional spacetime $\mathbb{R}^d$, with a metric $(-,+,+,\ldots, +)$. In M--theory $d=11$, for strings we have $d=10$. The $\xi^l$ are the longitudinal worldvolume coordinates on the D--branes that the transverse functions $x^j=x^j(\xi)$ depend on. Being matrices, the $x^j(\xi)$ are Lie--algebra valued,
\begin{equation}
x^j(\xi)=\sum_{a=1}^{n^2-1}x_a^j(\xi)T_a,
\label{barbonquetefollenhijoputa}
\end{equation}
where the $T^a$ generate $su(n)$ in the adjoint representation. To the Lagrangian density of eqn. (\ref{luisibanezquetefollenkabrondemierda}),
\begin{equation}
{\rm l}=-\frac{1}{2}{\rm tr}\,(D_lx^jD_lx^j)+\frac{1}{4}{\rm tr}\,\sum_{i\neq j}[x^i,x^j]^2,
\label{mekagoentodoslosdelaautonoma}
\end{equation}
there corresponds a Hamiltonian density
$$
{\rm h}=\frac{1}{2}{\rm tr}\,(p^jp^j)+\frac{1}{2}{\rm tr}\,(\partial_sx^j\partial_sx^j)+{\rm i}\,{\rm tr}\,(p^j[A_0,x^j])
$$
\begin{equation}
+{\rm i}\,{\rm tr}\,(\partial_sx^j[A_s,x^j])-\frac{1}{2}{\rm tr}\,[A_s, x^j]^2-\frac{1}{4}\sum_{i\neq j}{\rm tr}\,[x^i,x^j]^2,
\label{labastidamekagoentuputakara}
\end{equation}
where the subindex $s$ stands for the spacelike, longitudinal coordinates $\xi^1, \ldots, \xi^p$, and the (adjoint--valued) $p^j$ are the canonical momenta conjugate to the $x^j$.

\section{Dynamics of the Higgs sector}\label{kabronbarbonmarikon}

\subsection{Preliminaries}\label{kabronbarbon}

The equal--time canonical Poisson brackets (CPB) between coordinates and momenta for the field theory described by the action (\ref{luisibanezquetefollenkabrondemierda}) read
\begin{equation}
\left\{x^i_a(\xi^0,\xi), p^k_b(\xi^0, \xi')\right\}_{\rm CPB}=\delta_{ab}\delta^{ik}\delta^{(p}(\xi-\xi'),
\label{mekagoentuputakararamallo}
\end{equation}
where $\delta^{(p}(\xi-\xi')$ refers to the $p$ spacelike, longitudinal coordinates along the D$p$--brane. This delta function disappears when $p=0$, in which case the CPB (\ref{mekagoentuputakararamallo}) simplify to
\begin{equation}
\left\{x^i_a(\xi^0), p^k_b(\xi^0)\right\}_{\rm CPB}=\delta_{ab}\delta^{ik}.
\label{mekagoentuputakaralabastidademierda}
\end{equation}
As dictated by the M(atrix) model of M--theory \cite{BFSS}, in what follows we will set $p=0$ in the action (\ref{luisibanezquetefollenkabrondemierda}), so the corresponding CPB are given by eqn.  (\ref{mekagoentuputakaralabastidademierda}). The classical phase space ${\cal M}$ of the Higgs sector has the $2d(n^2-1)$ Darboux coordinates $x^{i}_a$ and $p^{k}_b$: there are $d$ transverse coordinates to a D$0$--brane, all of which are $su(n)$--valued. Eventually we will let $n\to\infty$.

At this point we pause to introduce some notation. The classical phase space ${\cal M}$ of the Higgs sector is has classical dynamics governed by a time--independent Hamiltonian function $H$. As such ${\cal M}$ will be a Poisson manifold, with $C^{\infty}({\cal M})$ as its algebra of smooth functions. Eqn. (\ref{mekagoentuputakaralabastidademierda}) gives its canonical Poisson brackets. Let us consider the 
matrix variables
\begin{equation}
x^{i}:=x^{i}_aT_a,\qquad p^{k}:=p^{k}_aT_a,\qquad x^{i}_a, p^{k}_b\in C^{\infty}({\cal M}),
\label{barbonketepartaunrayo}
\end{equation}
where the $T_a$ generate the adjoint representation of an arbitrary simple, real, finite--dimensional, compact Lie algebra $\mathfrak{g}$. Now $\mathfrak{g}$ supports Lie brackets
\begin{equation}
[\cdot\,,\cdot]\colon \mathfrak{g}\times \mathfrak{g}\longrightarrow \mathfrak{g}
\label{ramallocabron}
\end{equation}
which, in the basis $T_a$, read
\begin{equation}
[T_a,T_b]=\omega_{ab}^{\;\; c}T_c.
\label{mierdaparavr}
\end{equation}
The universal enveloping algebra ${\cal U}(\mathfrak{g})$ contains the quadratic Casimir operator $k^{ab}T_aT_b$, where $k^{ab}$ is the inverse of the Killing metric 
\begin{equation}
k_{ab}=\omega_{ac}^{\;\; d}\omega_{bd}^{\;\; c}.
\label{ramallomelapagaras}
\end{equation}
In the adjoint representation we have $k^{ab}T_aT_b=c_A{\bf 1}$.

The phase space ${\cal M}$ of the Higgs sector is noncompact, both for finite and infinite $n$. Upon quantisation, let ${\cal H}$ denote the Hilbert space of quantum states. The latter will also be infinite--dimensional because ${\cal M}$ is noncompact. Classical functions on phase space $f\in C^{\infty}({\cal M})$ become quantum operators $F$ on Hilbert space. This we denote as $F\in{\cal O}({\cal H})$, where ${\cal O}({\cal H})$ stands for the algebra of observables. We use lowercase letters $f$ for classical functions and uppercase letters $F$ for their quantum counterparts; the only exception to this rule is the Hamiltonian, denoted $H$ both as a classical function and as a quantum operator. All functions and all operators will be time--independent. The algebra ${\cal C}^{\infty}({\cal M})$ supports classical Poisson brackets (CPB), {\it i.e.}, an antisymmetric, bilinear map
\begin{equation}
\left\{\cdot\,,\cdot\right\}_{\rm CPB}\colon {\cal C}^{\infty}({\cal M})\times {\cal C}^{\infty}({\cal M})\longrightarrow {\cal C}^{\infty}({\cal M})
\label{labastidaquetepartaunrayo}
\end{equation}
satisfying the Jacobi identity and the Leibniz derivation rule.
Upon quantisation, the algebra of functions $C^{\infty}({\cal M})$ on classical phase space gets replaced by the algebra of operators ${\cal O}({\cal H})$ on Hilbert space. 
The quantum Poisson bracket $\left[\cdot\,,\cdot\right]_{\rm QPB}$ is an antisymmetric, bilinear map
\begin{equation}
\left[\cdot\,,\cdot\right]_{\rm QPB}\colon {\cal O}({\cal H})\times {\cal O}({\cal H})\longrightarrow {\cal O}({\cal H})
\label{labastidacabronquetepartaunrayo}
\end{equation}
also satisfying the Jacobi identity and the Leibniz derivation rule.

Although in this letter we are interested in the superposition of an {\it infinite}\/ number of D0--branes, we first review in sections \ref{casposoramallo}, \ref{barbobchupamelapollakabron} the corresponding results for finite $n$ following ref. \cite{GOLM}.

\subsection{The classical theory}\label{casposoramallo}

In ref. \cite{GOLM} we have defined, for finite $n$, the tensor product algebra 
\begin{equation}
{\cal C}({\cal M}, \mathfrak{g}):=C^{\infty}({\cal M})\otimes {\cal U}(\mathfrak{g}).
\label{ramallochupameelcarallo}
\end{equation}
It qualifies as a Poisson algebra because one can endow it with Poisson brackets
\begin{equation}
\{\cdot\,, \cdot\}_{{\cal C}(\cal M,\mathfrak{g})}\colon {\cal C}({\cal M}, \mathfrak{g})\times {\cal C}({\cal M}, \mathfrak{g})\longrightarrow {\cal C}({\cal M}, \mathfrak{g}).
\label{avzrmekago}
\end{equation}
For this purpose we have first defined the canonical bracket $\{x^{i}, p^k\}_{{\cal C}({\cal M}, \mathfrak{g})}$. Then any other bracket follows by requiring antisymmetry, bilinearity and the Leibniz derivation rule. For a compact $\mathfrak{g}$ we can pick $k^{ab}=\delta^{ab}$. Using eqn. (\ref{mekagoentuputakaralabastidademierda}) we have set,
\begin{equation}
\{x^{i},p^k\}_{{\cal C}({\cal M}, \mathfrak{g})}:=\{x^{i}_a, p^k_{b}\}_{\rm CPB}\,T_aT_b=\delta^{ik}\delta_{ab}\,T_aT_b=\delta^{ik}\,T_aT_a,
\label{putonramallo}
\end{equation}
{\it i.e.},
\begin{equation}
\{x^{i},p^k\}_{{\cal C}({\cal M}, \mathfrak{g})}=\delta^{ik}c_A{\bf 1}.
\label{ramallomekagoentuputakara}
\end{equation}
Then the Jacobi identity holds, and ${\cal C}({\cal M}, \mathfrak{g})$ becomes a Poisson algebra. The underlying $\mathfrak{g}$ appears through its quadratic Casimir eigenvalue $c_A$. In a representation $R$ other than the adjoint, $c_A$ is replaced with the corresponding quadratic Casimir eigenvalue $c_R$. The right--hand side of (\ref{ramallomekagoentuputakara}) is central within ${\cal C}({\cal M}, \mathfrak{g})$ as it should. For $\mathfrak{g}=su(n)$ we have $c_A=n$, and the fundamental brackets (\ref{ramallomekagoentuputakara}) read
\begin{equation}
\{x^{i},p^k\}_{{\cal C}({\cal M}, su(n))}=\delta^{ik}n{\bf 1}.
\label{ramallohijoputamekagoentuputakara}
\end{equation}
{}Finally the time evolution of an arbitrary $f\in {\cal C}({\cal M}, \mathfrak{g})$ is given by
\begin{equation}
\frac{{\rm d}f}{{\rm d}t}=\{f, H\}_{{\cal C}({\cal M}, \mathfrak{g})}.
\label{kap}
\end{equation}

\subsection{The quantum theory}\label{barbobchupamelapollakabron}

Still following ref. \cite{GOLM}, where $n$ is finite, we have quantised the dynamics of section \ref{kabronbarbonmarikon} by first assuming that we {\it turn off}\/ the Lie--algebra degrees of freedom. This has been achieved by separating all $n$ branes from each other, so no two of them remain coincident \cite{WITTEN}. Then $su(n)$ breaks into $n-1$ copies of $u(1)$. Effectively we are left with $n-1$ independent copies of $C^{\infty}({\cal M})$, placed along the diagonal of an $(n^2-1)\times (n^2-1)$ matrix. Now $C^{\infty}({\cal M})$ can be quantised by standard methods to yield the algebra ${\cal O}({\cal H})$ of quantum observables on Hilbert space ${\cal H}$. After this operation we let all $n$ branes coincide again, and we consider the tensor product algebra
\begin{equation}
{\cal Q}({\cal H}, \mathfrak{g}):={\cal O}({\cal H})\otimes {\cal U}(\mathfrak{g}),
\label{ramalloeresunkasposdemierda}
\end{equation}
which is the quantum analogue of the classical algebra (\ref{ramallochupameelcarallo}).
Now ${\cal Q}({\cal H}, \mathfrak{g})$ will qualify as an algebra of quantum operators if we can endow it with quantum Poisson brackets
\begin{equation}
[\cdot\,,\cdot]_{{\cal Q}({\cal H}, \mathfrak{g})}\colon
{\cal Q}({\cal H}, \mathfrak{g})\times {\cal Q}({\cal H}, \mathfrak{g})\longrightarrow {\cal Q}({\cal H}, \mathfrak{g}).
\label{labastidamarikononquetepartaunrayo}
\end{equation}
{}For this it suffices to define $[X^{i},P^k]_{{\cal Q}({\cal H}, \mathfrak{g})}$ as the quantum analogue of eqn. (\ref{ramallomekagoentuputakara}),
\begin{equation}
[X^{i},P^k]_{{\cal Q}({\cal H}, \mathfrak{g})}={\rm i}\hbar\delta^{ik}c_A{\bf 1}.
\label{ramallomekagoentuputamama}
\end{equation}
Then ${\cal Q}({\cal H}, \mathfrak{g})$ qualifies as a Poisson algebra. That is, extending the brackets (\ref{ramallomekagoentuputamama}) to all ${\cal Q}({\cal H}, \mathfrak{g})$ by requiring linearity, antisymmetry and the Leibniz derivation rule automatically ensures that the Jacobi identity is satisfied. Finally the time evolution of an arbitrary $F\in {\cal Q}({\cal H}, \mathfrak{g})$ is given by the quantum counterpart of eqn. (\ref{kap}),
\begin{equation}
{\rm i}\hbar\frac{{\rm d}F}{{\rm d}t}=[F, H]_{{\cal Q}({\cal H}, \mathfrak{g})}.
\label{kappp}
\end{equation}

\subsection{The limit $n\to\infty$}\label{casasguarro}

Next we take the limit $n\to\infty$ in sections \ref{casposoramallo}, \ref{barbobchupamelapollakabron}. This limit must be taken with care, as the right--hand sides of eqns. (\ref{ramallomekagoentuputakara}), (\ref{ramallomekagoentuputamama}) diverge when $n\to\infty$.
Let an orthonormal basis of infinite--dimensional, complex, separable Hilbert space ${\cal H}$ be given. Let $E_{kl}$ be the operator whose matrix reads, in this basis, 
\begin{equation}
(E_{kl})_{\alpha\beta}:=\delta_{k\alpha}\delta_{l\beta}, \qquad k,l,\alpha,\beta\in\mathbb{Z}^+.
\label{coletita}
\end{equation}
The $E_{kl}$ satisfy the algebra 
\begin{equation}
[E_{kl},E_{mq}]=\delta_{lm}E_{kq}-\delta_{qk}E_{ml}.
\label{konm}
\end{equation}
Let $U({\infty})$ denote the inductive limit of the unitary group $U(n)$ and $u({\infty})$ its Lie algebra. This inductive limit is the gauge group (or algebra) that one obtains upon stacking together an infinite number of branes. From now on we set $\mathfrak{g}=u({\infty})$. The latter is the Lie algebra over $\mathbb{R}$ generated by all selfadjoint operators with only a finite number of nonzero entries. A basis of $u({\infty})$ is spanned by 
\begin{equation}
\sigma_{kl}:=E_{kl}+E_{lk},\qquad
\tau_{kl}:=-{\rm i}E_{kl}+{\rm i}E_{lk},\qquad
\xi_{kk}:=E_{kk},
\label{toninsimpatico}
\end{equation}
and an arbitrary element $u\in u({\infty})$ is given by
\begin{equation}
u=\sum_{kl}x^{kl}\sigma_{kl}+\sum_{kl}y^{kl}\tau_{kl}+\sum_{k}z^{kk}\xi_{kk},
\label{toninbello}
\end{equation}
where $x^{kl}, y^{kl}, z^{kk}$ are real numbers, only finitely many of them nonzero. The Lie algebra $u({\infty})$ reads in this basis 
\begin{eqnarray}
[\sigma_{kl},\sigma_{mq}]&=&{\rm i}\left(\delta_{lm}\tau_{kq}+\delta_{kq}\tau_{lm}+\delta_{lq}\tau_{km}+\delta_{km}\tau_{lq}\right)\cr
[\sigma_{kl},\tau_{mq}]&=&{\rm i}\left(\delta_{lq}\sigma_{km}-\delta_{lm}\sigma_{kq}+\delta_{kq}\sigma_{ml}-\delta_{km}\sigma_{ql}\right)\cr
[\sigma_{kl},\xi_{mm}]&=&{\rm i}\left(\delta_{lm}\tau_{km}+\delta_{km}\tau_{lm}\right)\cr
[\tau_{kl},\tau_{mq}]&=&{\rm i}\left(\delta_{lm}\tau_{qk}+\delta_{lq}\tau_{km}+\delta_{kq}\tau_{ml}+\delta_{km}\tau_{lq}\right)\cr
[\tau_{kl},\xi_{mm}]&=&{\rm i}\left(\delta_{km}\sigma_{ml}-\delta_{lm}\sigma_{mk}\right)\cr
[\xi_{mm},\xi_{qq}]&=&0.
\label{toninmierda}
\end{eqnarray}
Two important points should now be observed.

{}First, we regularise the divergence in $c_A(n)$ as $n\to\infty$ by replacing it with its regularised value $c_A^{\rm reg}(n):=n^{-1}c_A(n)$ and then taking the limit $n\to\infty$. Thus we have $\lim_{n\to\infty}c_A^{\rm reg}(n)=1$. Our regularisation prescription tells us that the {\it regularised}\/ quadratic Casimir eigenvalue $c_A^{\rm reg}$ counts not the number of branes in the stack, which is infinite, but rather the {\it density}\/ of branes in the stack. Alternatively, but equivalently, using the structure constants given in eqn. (\ref{toninmierda}) one may compute the Killing metric (\ref{ramallomelapagaras}) for $u({\infty})$. In so doing one comes across terms containing $\delta_{kk}$. When $n$ is finite one has $\delta_{kk}=n$; this is in fact the origin of $c_A=n$. When $n\to\infty$, the definition (\ref{ramallomelapagaras}) for the  Killing metric diverges. However there is one physically meaningful regularisation, namely, to keep all terms containing $\delta_{kk}$, and to factor them out at the very end. This is equivalent to our previous regularisation prescription. These conclusions are not altered by the fact that, for finite $n$, one should actually use a basis of traceless operators rather than those given in eqn. (\ref{toninsimpatico}); the corresponding structure constants may be obtained from eqn. (\ref{toninmierda}).

Second, in the limit $n\to\infty$, the algebra of observables ${\cal O}({\cal H})$ equals the universal enveloping algebra ${\cal U}(\mathfrak{g})$ of the unitary Lie algebra $\mathfrak{g}=u({\cal H})$,
\begin{equation}
{\cal O}({\cal H})={\cal U}(u({\cal H})).
\label{mekagoentuputakarabarbondemierda}
\end{equation}
As might have been guessed by now, the unitary Lie algebra $u({\cal H})$ on  ${\cal H}$ is {\it not}\/ to be confused with the inductive limit $u(\infty)$. While all operators of the form (\ref{toninbello}) belong to $u({\cal H})$, and thus 
\begin{equation}
u({\infty})\subset u({\cal H}), 
\label{kakapsoe}
\end{equation}
the latter contains selfadjoint operators that cannot be written as a sum (\ref{toninbello}), {\it i.e.}, as a sum with only {\it finitely many}\/ nonzero entries. An arbitrary element of $u({\cal H})$ can be obtained as a sum (\ref{toninbello}) if we allow for an {\it infinite}\/ number of nonzero $x^{kl}$, $y^{kl}$, $z^{kk}$. Thus the Lie algebra $u({\cal H})$ obtained as a result of quantisation is larger than the Lie algebra $u(\infty)$ obtained as the gauge symmetry on an infinite number of coincident branes. This is in nice agreement with a corresponding fact observed in ref. \cite{GOLM}, namely, that one must distinguish between {\it nonabelianity}\/ and {\it noncommutativity}. The latter arises upon quantisation, being due to the fact that $\hbar\neq 0$; the former is the result of stacking together more than one brane, so the gauge group is nonabelian.
As such, nonabelianity exists already before, and independently of, quantisation.

We can round up our discussion by stating that, in the limit $n\to\infty$, eqn. (\ref{ramalloeresunkasposdemierda}) becomes
\begin{equation}
{\cal Q}({\cal H}, {u({\infty})}):={\cal O}({\cal H})\otimes{\cal U}(u({\infty}))={\cal U}(u({\cal H}))\otimes{\cal U}(u({\infty})).
\label{alfonsoramalloeresunkasposdemierda}
\end{equation}
With our regularisation one has to replace eqn. (\ref{ramallomekagoentuputamama}) with 
\begin{equation}
[X^{i},P^k]_{{\cal Q}({\cal H}, u({\infty}))}={\rm i}\hbar\delta^{ik}{\bf 1},
\label{ramallomekagoentumama}
\end{equation}
and ${\cal Q}({\cal H}, u({\infty}))$ qualifies as a Poisson algebra.

\section{Constructing noncommutative manifolds}\label{kabronramallo}

When do the algebras (\ref{ramallochupameelcarallo}) and (\ref{alfonsoramalloeresunkasposdemierda}) give rise to noncommutative manifolds? Whenever the corresponding algebra qualifies as a $C^{\star}$--algebra, it has an interpretation as a manifold \cite{NCG, REV}, eventually noncommutative. Let us recall that a sufficient condition for the algebra of continuous functions $C^0(V)$ on a topological space $V$ to be a $C^{\star}$--algebra is that $V$ be compact \cite{NCG}. This is not the case of ${\cal M}$, which is noncompact. On the other hand, the universal enveloping algebra ${\cal U}(\mathfrak{g})$ is a $C^{\star}$--algebra when $\mathfrak{g}$ is finite--dimensional, which is also not the case of $u({\cal H})$ nor that of $u(\infty)$. Can we construct any noncommutative manifold with the ingredients at hand?

Let us analyse the quantum case first. The infinite--dimensional Lie algebra $u({\cal H})$ can still give rise to a $C^{\star}$--algebra if we restrict ourselves to $u_B({\cal H})$, {\it i.e.}, to the subalgebra of {\it bounded}, selfadjoint operators \cite{THIRRING}. Then the universal enveloping algebra ${\cal U}(u_B({\cal H}))$ qualifies as a $C^{\star}$--algebra, and so does ${\cal U}(u(\infty))$ too. Deferring for a while issues concerning the definition of a norm on a tensor product algebra, as well as its completeness, a $C^{\star}$--algebra is obtained from eqn. (\ref{alfonsoramalloeresunkasposdemierda}) upon setting
\begin{equation}
{\cal Q}({\cal H}, {u_B({\cal H})}):={\cal U}(u_B({\cal H}))\otimes {\cal U}(u(\infty)).
\label{kakaxavr}
\end{equation}

The corresponding classical case, eqn. (\ref{ramallochupameelcarallo}), requires some care because  $C^{\infty}({\cal M})$ does not qualify as a $C^{\star}$--algebra, not even when ${\cal M}$ is compact \cite{THIRRING}.  Now the algebra of {\it continuous}\/ functions on a {\it compact}\/ manifold does qualify as a $C^{\star}$--algebra \cite{THIRRING}, but ${\cal M}$ is noncompact. Let us embed a certain smooth manifold ${\cal K}$ within ${\cal M}$, so that (some of) the $x^i$, $p^k$ are coordinates on ${\cal K}$. The latter need not be compact; eventually ${\cal K}$ may coincide with ${\cal M}$ itself. Let us consider the algebra $C_b^0({\cal K})$ of complex--valued, {\it continuous, bounded}\/ functions on ${\cal K}$. Now $C_b^0({\cal K})$ qualifies as a $C^{\star}$--algebra \cite{THIRRING}. Depending on the amount of supersymmetry one wishes to preserve, typical examples for ${\cal K}$ could be a Riemann surface, a K3, or a Calabi--Yau manifold, among others. We define the tensor product
\begin{equation} 
{\cal C}({\cal K}, u({\infty})):=C^{0}_b({\cal K})\otimes {\cal U}(u({\infty})).
\label{marikonramallochupameelcarajo}
\end{equation}
Again deferring for a while issues concerning the definition of a norm on a tensor product algebra, as well as its completeness,
eqn. (\ref{marikonramallochupameelcarajo}) defines a $C^{\star}$--algebra that one can naturally associate with a given smooth manifold ${\cal K}$, by just placing it orthogonally to a stack of an infinite number of coincident D0--branes. In this sense we may call ${\cal C}({\cal K}, u({\infty}))$ {\it the}\/ nonabelian partner of ${\cal K}$, and ${\cal Q}({\cal H}, {u_B({\cal H})})$ in eqn. (\ref{kakaxavr}) {\it the}\/ nonabelian, noncommutative partner of ${\cal M}$. The latter enters the $C^{\star}$--algebra (\ref{kakaxavr}) after quantising ${\cal M}$ into ${\cal H}$, although this is not explicitly reflected in our notation. A more precise, though lengthier, notation would be ${\cal Q}({\cal H}({\cal M}), {u_B({\cal H})})$. 

As a consistency check on eqn. (\ref{marikonramallochupameelcarajo}), let us separate all D0--branes from each other, so that no two of them remain coincident. This causes $u({\infty})$ to break to an (infinite) product of $u(1)$'s. Effectively we are placing one copy of $C^{0}_b({\cal K})$ in each diagonal entry of an infinite--dimensional matrix, all offdiagonal entries being zero. The multiplication law has become commutative, and we are left with the algebra of continuous, bounded functions on ${\cal K}$. When the latter is compact this is as good as knowledge of ${\cal K}$ itself. Other breaking patterns are also possible. Thus keeping $1<N<\infty$ branes coincident and separating out all the rest from each other, as well as from the previous $N$ branes, we are left with two sectors. The one containing an infinite number of separate branes is commutative as described above. The sector containing $1<N<\infty$ coincident branes is nonabelian, and it corresponds to the the situation described in ref. \cite{GOLM}.

We close this section with some remarks concerning the definition of a $C^{\star}$--structure on a tensor product of $C^{\star}$--algebras. For details see, {\it e.g.}, ref. \cite{STAR}. Given two $C^{\star}$--algebras $A$ and $B$, with respective norms $\vert\vert\cdot\vert\vert_A$ and 
$\vert\vert\cdot\vert\vert_B$, let us form their algebraic tensor product $A\otimes B$. A {\it cross--norm}\/ on $A\otimes B$ is a norm that verifies $\vert\vert a\otimes b\vert\vert=\vert\vert a\vert\vert_A\cdot\vert\vert b\vert\vert_B$ for all $a\in A$ and all $b\in B$. One shows that on $A\otimes B$ there exists a cross--norm, denoted $\vert\vert\cdot\vert\vert_{\rm min}$, that defines a $C^{\star}$--norm on the tensor product $A\otimes B$. The latter can be completed with respect to $\vert\vert\cdot\vert\vert_{\rm min}$ in order to render $A\otimes B$ a complete space and thus a {\it bona fide}\/ $C^{\star}$--algebra, called the $C^{\star}$--tensor product of $A$ and $B$. Our eqns. (\ref{kakaxavr}), (\ref{marikonramallochupameelcarajo}) are to be understood in this sense.

\section{Conclusions}\label{ramallotienesmuchacaspa}

Superimposed D--branes have matrix--valued functions as their transverse coordinates, since the latter take values in the Lie algebra of the gauge group inside the stack of coincident branes. This leads to considering a classical dynamics where the multiplication law for coordinates and/or momenta, being given by matrix multiplication, is nonabelian. Quantisation further introduces noncommutativity as a deformation in powers of Planck's constant $\hbar$. In this letter we have presented a framework to describe the classical and quantum dynamics of Lie--algebra valued coordinates and momenta. The corresponding classical action is given by eqn. (\ref{luisibanezquetefollenkabrondemierda}) with $p=0$. This is the action of the Higgs sector in the low--energy theory living inside the stack of $n$ coincident D$0$--branes, whose {\it transverse}\/ excitations it describes. This property is interpreted in ref. \cite{WITTEN} as meaning that the coordinates orthogonal to the coincident branes, once quantised,  describe quantum fluctuations of the branes themselves. Eventually we let $n\to\infty$, in order to complement the results of ref. \cite{GOLM}, where $n$ was always taken to be finite. Along the way to setting $n=\infty$, a field--theoretic regularisation has been performed that was absent for all finite $n$. This is a remnant of the fact that, although the {\it transverse}\/ directions to a stack of D--branes are at most 10, we are actually dealing with an infinite number of degrees of freedom. Our regularisation amounts to replacing the number of branes, which is infinite, with the density of branes in the stack.

Our main results are collected in eqns. (\ref{kakaxavr}) and (\ref{marikonramallochupameelcarajo}). Eqn. (\ref{marikonramallochupameelcarajo}) describes the construction of a nonabelian $C^{\star}$--algebra that one can naturally associate with a given smooth manifold ${\cal K}$ by embedding it within ${\cal M}$, the classical phase space of the Higgs sector. As such, ${\cal K}$ lies orthogonally to an infinite number of coincident D0--branes. The amount of conserved supersymmetry in eqns. (\ref{kakaxavr}) and (\ref{marikonramallochupameelcarajo}) is the complement of the one in which the branes are taken to wrap around ${\cal K}$. This situation is complementary to the one considered in ref. \cite{NOI}, when a noncommutative Riemann surface arose by wrapping the branes on a Riemann surface and turning on a background $B$--field \cite{CDS}. Instead, here the D$p$--branes wrap a Minkowskian $\mathbb{R}^{p+1}$ in the absence of a Neveu--Schwarz $B$--field and are orthogonal to a certain manifold ${\cal K}$. Our approach of stacking branes orthogonally to a given manifold ${\cal K}$, rather than wrapping them around ${\cal K}$, improves on the construction of ref. \cite{NOI}. By placing the branes {\it orthogonally}\/ to ${\cal K}$ rather than {\it wrapping}\/ ${\cal K}$, one gets around the need of cancelling anomalies \cite{ANOMALY}. The anomaly--cancellation condition places severe constraints on the topology of the manifold ${\cal K}$ being wrapped and  on the magnetic fluxes across it. Finally our method requires no knowledge of the fundamental group of ${\cal K}$.

{\bf Acknowledgements}

It is a great pleasure to thank J. de Azc\'arraga for encouragement and support. The author thanks Albert-Einstein-Institut f\"ur Gravitationsphysik (Potsdam, Germany), where this work was begun, for hospitality. This work has been partially supported by research grant BFM2002--03681 from Ministerio de Ciencia y Tecnolog\'{\i}a, by research grant GV2004-B-226 from Generalitat Valenciana, by EU FEDER funds, by Fundaci\'on Marina Bueno and by Deutsche Forschungsgemeinschaft.

\end{document}